\documentclass{article}
\usepackage{spconf,amsmath,graphicx}
\usepackage{amssymb}
\usepackage{hyperref}
\usepackage{url}
\usepackage{multirow}
\usepackage{verbatim}

\DeclareMathOperator{\EX}{\mathbb{E}}
\newcommand{\argmax}{\mathop{\rm max}\limits}
\newcommand{\argmin}{\mathop{\rm min}\limits}

\title{High-quality nonparallel voice conversion based on \\cycle-consistent adversarial network}
\name{Fuming Fang$^{1}$, Junichi Yamagishi$^{1,2}$, Isao Echizen$^{1}$, Jaime Lorenzo-Trueba$^{1}$
\sthanks{This work was partially supported by MEXT KAKENHI Grant Numbers 15H01686, 16H06302, 17H04687.}
\sthanks{A demonstration of audio samples is available at \url{https://fangfm.github.io/icassp2018.html}}}
\address{$^1$National Institute of Informatics, Japan \\
  $^2$University of Edinburgh, UK \\
 {\small \tt \{fang, jyamagis, iechizen, jaime\}@nii.ac.jp}}
\begin{document}
\ninept
\maketitle
\begin{abstract}
Although voice conversion (VC) algorithms have achieved remarkable success along with the development of machine learning, superior performance is still difficult to achieve when using nonparallel data. In this paper, we propose using a cycle-consistent adversarial network (CycleGAN) for nonparallel data-based VC training. A CycleGAN is a generative adversarial network (GAN) originally developed for unpaired image-to-image translation. A subjective evaluation of inter-gender conversion demonstrated that the proposed method significantly outperformed a method based on the Merlin open source neural network speech synthesis system (a parallel VC system adapted for our setup) and a GAN-based parallel VC system. This is the first research to show that the performance of a nonparallel VC method can exceed that of state-of-the-art parallel VC methods.
\end{abstract}
\section{Introduction}
\label{sec:intro}
Voice conversion (VC) is a technique for modifying the speech signals of a source speaker to match those of a target speaker so that it sounds as if the target speaker had spoken while keeping the linguistic information unchanged~\cite{196671,vc}. A major application of VC is to personalize and create new voices for text-to-speech (TTS) synthesis systems~\cite{vc_overview}. Other applications include speaking aid devices that generate more intelligible voice sounds to help people with speech disorders~\cite{KAIN2007743}, movie dubbing~\cite{dubbing}, language learning~\cite{language_learning}, singing voice conversion~\cite{singingconversion}, and games.

The goal of VC is to find a mapping between the source and target speakers' speech features. Vector quantization (VQ), a Gaussian mixture model (GMM), or an artificial neural network (ANN) can be used as a mapping function or as a modeling framework~\cite{vq_vc, gmm_vc_original, ann_vc}. Since their parameters must be learned from a database, they are corpus-based techniques.
Depending on whether the training data obtained from the source and target speakers consists of repetitions of the same linguistic contents or not, VC can be categorized into {\it parallel} and {\it nonparallel} systems. In parallel systems, the training data for both speakers consists of the same linguistic content and thus forms a parallel corpus. Since the acoustic features of the source and target speaker that are similar will be closely related, they can be easily aligned, facilitating estimation of the mapping model parameters. As a result, parallel systems typically show high performance.

In nonparallel systems, the training data consists of different linguistic content and thus forms a nonparallel corpus. Since linguistic features are not shared, automatically matching the acoustic features of the two speakers that are similar is more difficult. As a result, the mapping model is harder to train, and performance is typically worse than that of parallel systems. However, since any utterance spoken by either speaker can be used as a training sample, if a nonparallel VC system can achieve comparable performance, it will be more flexible, more practical, and more valuable than parallel VC systems. This is because nonparallel training data (no need for uttering the same sentence set) can be easily collected from a variety of sources such as YouTube videos. Moreover, it is impossible to build a parallel data set if the source and target speakers speak different languages or have different accents.

A potential way to improve the performance of nonparallel VC systems is to use a cycle-consistent adversarial network (CycleGAN)~\cite{CycleGAN}. A CycleGAN is a type of generative adversarial network (GAN)~\cite{gan} originally developed for unpaired image-to-image translation. The basic idea of a CycleGAN is that there exists an underlying relationship between distributions, so a cycle-consistency loss can be introduced to constrain part of the input information so that it is invariant when processed throughout the network while adversarial loss is used to make the distribution of the generated data and that of the real target data indistinguishable. As a result, the relationship between distributions can be learned using unpaired data without directly matching similar features. Previous work~\cite{CycleGAN} using this method demonstrated that zebras in a photograph could be converted into horses, winter into summer, and so on.

We have proposed a method that uses a CycleGAN to improve the performance of nonparallel VC systems. When a CycleGAN-based VC is being trained, each discriminator of the CycleGAN can be thought of as a judge who distinguishes whether an input is from a source speaker or from the target speaker. At the same time, its generators strive to confuse the discriminator while maintaining the linguistic information of the source speaker. This competition enables the generators to convert the speech of a speaker into that of another speaker. Subjective experiments demonstrated the effectiveness of the proposed method.

The rest of this paper is organized as follows. Section~\ref{sec:related_work} explains differences between the proposed method and previous ones. Section~\ref{sec:cycleGAN} gives a brief explanation of a CycleGAN. Section~\ref{sec:proposed_method} describes CycleGAN-based nonparallel VC. Sections~\ref{sec:setup} and~\ref{sec:results} present the experimental setup and results, respectively. Section~\ref{sec:limitation} discusses the results and analyzes some limitations of the proposed method. Finally Section~\ref{sec:conclusion} summarizes the key points and mentions future work.

\section{Related work}
\label{sec:related_work}
In this section, we discuss the differences between the proposed nonparallel VC method and several related parallel and nonparallel VC methods.

\subsection{Related parallel VC methods}
Among the related parallel VC methods, the one proposed by Stylianou et al.~\cite{gmm_vc_original} uses a GMM as the mapping model, in which the features of the source and target speakers that are similar are paired using a joint vector that represents the relationships between the two speakers. It is used by the GMM for parameter training. Toda et al.~\cite{gmm_vc} improved this GMM-based method by incorporating the consideration of dynamic features and global variance. Desai et al.~\cite{nn_vc} used a feed forward neural network (NN) as the mapping model, in which the features that are similar are paired and serve as input and supervisor signals for parameter training. To capture more context, Sun et al.~\cite{blstm_vc} extended the feed forward NN to bidirectional long short-term memory (BLSTM)~\cite{blstm} and achieved better performance. GANs have recently been shown to be an effective training method and have been used for NN-based VC. Kaneko et al.~\cite{gan_vc_para} applied a GAN to sequence-to-sequence VC and demonstrated that the use of GAN-based training criteria outperforms the use of traditional mean squared error (MSE)-based training criteria.

In short, the previous parallel VC methods require that the features of the two speakers that are similar be aligned and paired for training of the mapping model. However, the alignment is not always true~\cite{vc_overview}, so new errors may be introduced. In contrast, our proposed method does not require parallel training data and does not require alignment.

\subsection{Related nonparallel VC methods}
A number of nonparallel VC methods have been developed, and they can be roughly split into two types: feature-pair searching and individuality replacement. The feature-pair searching methods match the feature pairs of the source and target speakers that are similar and thus can learn a conversion model using a parallel training method. For example, Ye and Young~\cite{ye2004voice} used a hidden Markov model (HMM)-based speech recognizer to gather phone information on the basis of a given or recognized transcription. They then matched the pairs of similar features by using HMM state indices. There are also feature-pair-based methods that do not rely on phonetic or linguistic information, such as INCA, presented by Erro et al.~\cite{erro2010inca}. Their method iteratively looks for nearest neighbor feature pairs between the source and target speaker while also iteratively updating the conversion model to progressively improve matching to the target speaker. By incorporating the consideration of context and both source-to-target and target-to-source conversion during iterative search, Benisty et al.~\cite{tcinca} achieved further improvement.

The individuality replacement methods are based on the assumption that a segment of speech can be split into linguistic and speaker identity components so as to achieve conversion by replacing the speaker identity component. To represent speaker identity, Song et al.~\cite{song2013non} adapted a GMM from a pre-prepared background model using a maximum a posteriori (MAP) approach. Nakashika et al.~\cite{nakashika_vc} proposed a more accurate method in which an adaptive restricted Boltzmann machine uses weights composed of both common weights and speaker identity weights. These weights can be estimated from data obtained from multiple speakers. Hsu et al.~\cite{cvae_wgan_vc} proposed a replacement method in which a conditional variational autoencoder (C-VAE) and Wasserstein GAN (W-GAN)~\cite{wgan} are combined. The encoder of the C-VAE is used to generate a phonetic distribution while the decoder generates the target speech features by combining the distribution and speaker identity. The W-GAN distinguishes whether an input is from the target speaker or not.

Compared to these previous nonparallel VC methods, our proposed method is more straightforward. In that sense, the method of Hsu et al. is the most similar to ours as it uses a GAN to generate features similar to those of the target. Our method differs in that it does not split the linguistic information from the source speaker. Instead, part of the linguistic information is assumed to be invariant when processed throughout the network.

\section{Cycle-consistent adversarial network}
\label{sec:cycleGAN}
\begin{figure}[tb]
  \begin{center}
    \includegraphics[width=0.33\textwidth]{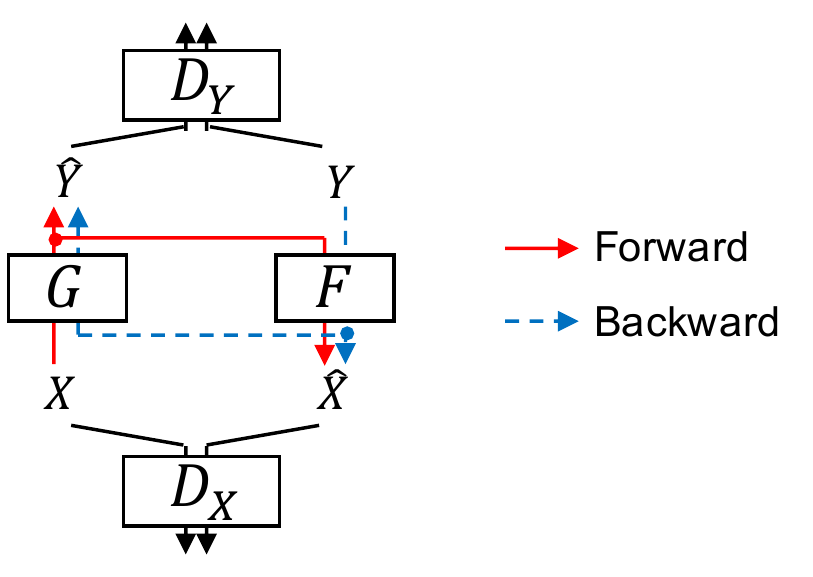}
    \vspace{-5mm}
    \caption{Diagram of a CycleGAN. $G$ and $F$ are generators; $D_X$ and $D_Y$ are discriminators. $X$ and $Y$ are the real distributions, and $\hat{Y}$ and $\hat{X}$ represent the corresponding generated distributions, respectively.}
    \label{fig:cyclegan}
  \end{center}
  \vspace{-5mm}
\end{figure}

A CycleGAN consists of two generators ($G$ and $F$) and two discriminators ($D_X$ and $D_Y$), as shown in Figure~\ref{fig:cyclegan}. Generator $G$ serves as a mapping function from distribution $X$ to distribution $Y$, and generator $F$ serves as a mapping function from $Y$ to $X$. The discriminators aim to distinguish between the real and generated distributions, i.e., $D_X$ distinguishes $X$ from $\hat{X} = F(Y)$, and $D_Y$ distinguishes $Y$ from $\hat{Y} = G(X)$. The goal of this model is to learn the mapping functions given training samples $\{x_i\}_{i=1}^N\in X$ and $\{y_j\}_{j=1}^M\in Y$. To this end, two types of loss are defined as optimization objectives: {\it adversarial loss} and {\it cycle-consistency loss}. The adversarial loss makes $X$ and $\hat{X}$ or $Y$ and $\hat{Y}$ as similar as possible while the cycle-consistency loss guarantees that an input $x_i$ (or $y_j$) can retain its original form after passing through the two generators. By combining these losses, a model can be learned from unpaired training samples, and the learned mappings are able to map an input $x_i$ (or $y_j$) to a desired output $y_j$ (or $x_i$). Note that there are two cycle mapping directions in this model: $X \rightarrow \hat{Y} \rightarrow \hat{X}$ and $Y \rightarrow \hat{X} \rightarrow \hat{Y}$. This means that the two mappings can be learned simultaneously. To distinguish between the directions, the former is defined as {\it forward} cycle consistency, and the latter is defined as {\it backward} cycle consistency. Details of the optimization objectives are described below.

For the adversarial loss, the objective function for mapping $G$ and the corresponding discriminator $D_Y$ is defined as
\begin{align}
\mathcal{L}_{GAN} (G, D_Y, X, Y) &=\EX_{y\sim p_{data}(y)}[\log D_Y(y)] \nonumber \\
&+\EX_{x\sim p_{data}(x)}[\log(1-D_Y(G(x)))],
\label{eq:adversarial}
\end{align}
where $\EX$ means expectation. Strictly speaking, the second term on the right has expectation with respect to not only $x$ but also latent variable $z$, but we omit $z$ from the formulation to simplify the notation. The objective function for $F$ and $D_X$ has a similar formulation: $\mathcal{L}_{GAN} (F, D_X, Y, X)$. During training, $G$ and $F$ try to minimize these two objective functions while at the same time $D_Y$ and $D_X$ try to maximize them. The cycle-consistent loss function is analogous to the objective function of an autoencoder, which minimizes the difference between the input and output to reconstruct the input from the output. Thus, the cycle-consistent loss is defined as
\begin{align}
\mathcal{L}_{cyc} (G, F) &=\EX_{x\sim p_{data}(x)}[\| F(G(x))-x \|_1] \nonumber \\
&+\EX_{y\sim p_{data}(y)}[\|G(F(y))-y \|_1],
\label{eq:cycloss}
\end{align}
where $\| \cdot \|_1$ means L1 norm.
The full objective function combines the adversarial and cycle-consistent losses:
\begin{align}
\mathcal{L} (G, F, D_X, D_Y) &= \mathcal{L}_{GAN} (G, D_Y, X, Y) \nonumber \\
&+\mathcal{L}_{GAN} (F, D_X, Y, X) \nonumber \\
&+ \lambda \mathcal{L}_{cyc} (G, F),
\label{eq:fullloss}
\end{align}
where $\lambda$ controls the relative importance of the two losses. Finally, the model parameters are estimated by solving the following equation using the back propagation algorithm.
\begin{align}
G^*, F^* = \text{arg} \argmin_{G, F} \argmax_{D_X, D_Y} \mathcal{L} (G, F, D_X, D_Y)
\label{eq:final}
\end{align}

In practice, since the least squares loss is more stable than the negative log likelihood when conducting back propagation, $\mathcal{L}_{GAN}$ can be rewritten as $\mathcal{L}_{LSGAN}$~\cite{mao2017least}, e.g.,
\begin{align}
\mathcal{L}_{LSGAN} (G, D_Y, X, Y) &=\EX_{y\sim p_{data}(y)}[(D_Y(y)-1)^2] \nonumber \\
&+\EX_{x\sim p_{data}(x)}[D_Y(G(x))^2].
\label{eq:lsgan}
\end{align}

\section{Nonparallel VC based on CycleGAN}
\label{sec:proposed_method}
\begin{figure}[tb]
  \begin{center}
    \includegraphics[width=0.33\textwidth]{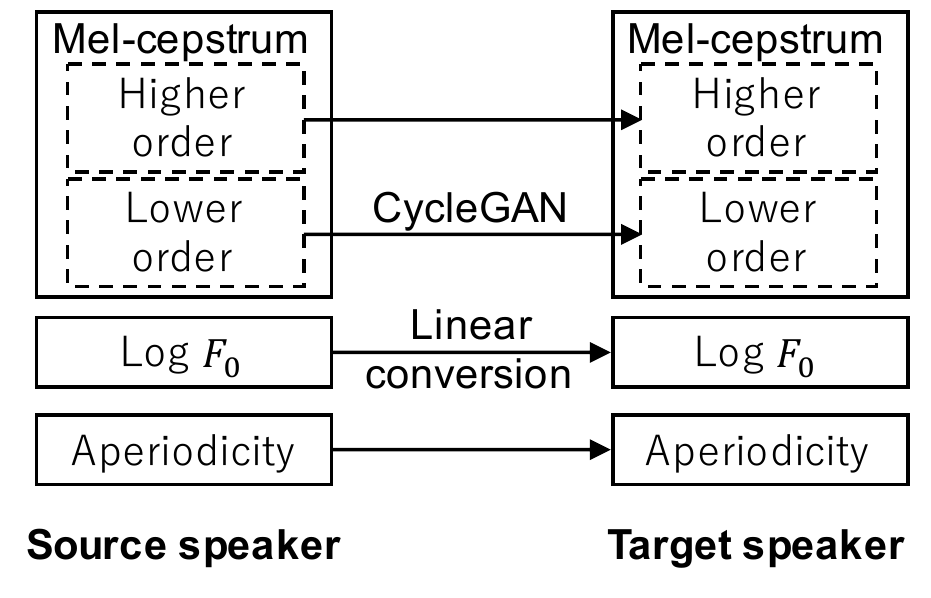}
    \vspace{-5mm}
    \caption{Overview of CycleGAN-based nonparallel voice conversion system.}
    \label{fig:framework}
  \end{center}
\vspace{-8mm}
\end{figure}

Figure~\ref{fig:framework} shows an overview of our CycleGAN-based nonparallel voice conversion system. Voice conversion is achieved by extracting, converting, and then synthesizing the speech features. The mel-cepstrum, fundamental frequency ($F_0$), and aperiodicity bands are the speech features used here. As shown in the figure, these components are converted separately. To facilitate mel-cepstrum conversion, it is first split into two sub-components: higher order and lower order. The former corresponds to the spectral fine structure, and the latter corresponds to the spectral envelope. We assume that the higher-order cepstral coefficients do not carry much speaker information since the corresponding parts of the mel-cepstrum always exhibit little change. Therefore, we directly copy these coefficients as part of the converted speech's features.
The lower-order cepstral coefficients are known to clearly reflect linguistic information and speaker identity. As such, we focus the efforts of the CycleGAN on the conversion of this particular component. For $F_0$ conversion, the source speaker's log $F_0$ is linearly transformed by equalizing the mean and the standard deviation of the target speaker's log $F_0$, a widely used method in the VC area~\cite{blstm_vc}. The aperiodicity component is directly copied when synthesizing the converted speech since it has no significant effect on the speaker characteristics of the synthesized speech~\cite{ap}.

For CycleGAN-based VC, $X$ and $Y$ correspond respectively to the distributions of the source and target speaker features (i.e., only the lower-order mel-cepstrum coefficients here). Therefore, training samples $\{x_i\}_{i=1}^N\in X$ and $\{y_j\}_{j=1}^M\in Y$ are collections of the mel-cepstral coefficients extracted from each frame of the source or target speaker's speech data included in a mini-batch. For each iteration of the back propagation, we randomly draw a mini-batch from a training dataset and compute Eq.~\ref{eq:final}.

\section{Experimental setup}
\label{sec:setup}
We compared the performance of our proposed CycleGAN-based nonparallel VC with those of two parallel VC methods (baselines) in terms of speech quality and speaker similarity by conducting a subjective evaluation. The first baseline method was based on the Merlin~\cite{wu2016merlin} open source neural network speech synthesis system from the University of Edinburgh. A part of its configuration was modified as described in subsection~\ref{ssec:systems} and other hyper-parameters were same as baseline of the Voice Conversion Challenge (VCC) 2016. With this setup, we achieved similar performance to the VCC2016 baseline. The second baseline method was a GAN~\cite{mao2017least}-based method, where the MSE criteria was additionally used to help training the model. All three methods performed inter-gender conversion, i.e., female-to-male and male-to-female conversions. The statistical significance analysis was based on an unpaired two-tail $t$-test with a 95\% confidence interval and Holm-Bonferroni compensation for the 3-way system comparison.

\subsection{Database and speech feature}
\label{ssec:data}
We used the ALAGIN Japanese Speech Database~\cite{alagin} Set B. This database contains data from ten speakers, but we used the data for only one male speaker (MTK) and one female speaker (FKN). There were ten sub-datasets (indexed A to J) for each speaker, and the corresponding utterance sets had the same index. Subsets A to D of the two speakers (i.e., 200 utterances/speaker) were used to create a parallel dataset for training of the baseline methods. Subsets A to D of the male speaker and subsets E to H of the female speaker were used to create a nonparallel dataset (i.e., 200 utterances/speaker) for the proposed method. We used subset I (50 utterances) for both the proposed and baseline methods for testing. Although the database contains transcriptions, we did not use them.

The audio data were sampled at 20 KHz with a bit depth of 16 bits. The mel-cepstrum, $F_0$, and aperiodicity bands were extracted using the WORLD~\cite{morise2016world} and speech signal processing toolkits (SPTK)~\cite{sptk2009speech}. The number of mel-cepstrum dimensions was set to 49: the first 25 were used as the lower order component, and the last 24 were used as the higher order component. To capture context, the first and second derivatives of the mel-cepstrum were used. As a result, 75-dimension feature vectors were created for learning the conversion models (i.e., 25 for each the statics, first derivative and second derivative). The features of the parallel datasets were aligned using dynamic time warping (DTW) while the nonparallel dataset did not undergo any matching pre-process.

\subsection{Network structure, training and conversion setup}
\label{ssec:systems}
The network structure of the Merlin-based baseline conversion model and generators as well as the discriminators of the GAN and the CycleGAN was a six-layer feed forward NN. The number of neurons in each hidden layer was 128, 256, 256, or 128. A sigmoid was used as the activation function for all hidden units. Both the GAN baseline and CycleGAN methods were implemented on the TensorFlow framework~\cite{tensorflow2015-whitepaper}. The default learning rate was set to 0.001 (0.0001 when updating the discriminators). Mini-batches were constructed from 128 randomly selected frames. The number of epochs was set to 60 for the Merlin baseline method and to 400 for the GAN and CycleGAN methods. The $\lambda$ in Eq.~\ref{eq:fullloss} was set to 10 when training the CycleGAN. Maximum likelihood parameter generation (MLPG)~\cite{tokuda2000speech} and post-filtering~\cite{YoshimuraTMKK05} were conducted to generate smooth speech parameters.

\subsection{Subjective evaluation setup}
A total of 300 ($=50~\text{utterances} \times 3~\text{methods} \times 2~\text{genders}$) converted utterances were compared with the corresponding natural reference utterance in terms of speech quality and speaker similarity. Both metrics were evaluated on a 1-to-5 Likert mean opinion score (MOS) scale. The evaluation was carried out by means of a crowdsourced web-based interface. The evaluators were first shown a web page on which they input their gender and age. They were then each asked to rate sets of 12 utterances randomly selected from the 300 utterances. They were limited to rating a maximum of six sets so that they would not become complacent about. Although they were able to play each sample utterance as many times as they wanted, they had to completely play the audio samples and answer all the questions displayed on the web page for their evaluations to be considered in the evaluation. A total of 110 evaluators produced a total of 7200 data points, which is equivalent to 24 evaluations per utterance.

\section{Experimental results}
\label{sec:results}
As shown in Table~\ref{tbl:subjectiveresults}, the proposed CycleGAN-based nonparallel VC method achieved significantly better performance than the parallel VC baseline methods in terms of both average speech quality and speaker similarity. This suggests that a nonparallel VC method with a CycleGAN can achieve performance superior to that of state-of-the-art parallel VC methods. 

We noticed that there was no improvement achieved by the proposed method in terms of male-to-female conversion compared to the GAN-based method. We also noticed that the male-to-female conversion had lower speaker similarity scores than the female-to-male one for all methods. One possible reason is $F_0$ mismatch due to using only the global mean and standard deviation of the target speaker's training data during conversion, and the female speaker's $F_0$ was highly variant in time. Another possible reason is the use of insufficient components of the mel-cepstrum (the first 25 dimensions) for conversion to a female voice. To further improve conversion performance, $F_0$ should be learned together with the conversion model, and the dimensions of the mel-cepstrum should be appropriately selected.

\begin{table}[tb]
  \centering
  \caption{Perceptual evaluation results on MOS scale for speech quality and speaker similarity. ``CycleGAN'' denotes proposed nonparallel VC method and ``GAN'' and ``Merlin-based baseline'' denote the two baseline methods based on parallel VC. ``F$\rightarrow$M'' and ``M$\rightarrow$F'' indicate female-to-male and male-to-female conversion, respectively.}
  \vspace{3pt}
  \label{tbl:subjectiveresults}
  \begin{tabular}{llcc}
  \hline
    \multicolumn{2}{c}{\textbf{METHOD}} & \textbf{QUALITY} & \textbf{SIMILARITY} \\ \hline
    CycleGAN & F$\rightarrow$M & 2.89 & 2.53 \\
    (Nonparallel VC) & M$\rightarrow$F & 2.50 & 1.76 \\
    & {\bf AVG.} & {\bf 2.69} & {\bf 2.15} \\ 
    \hline
    GAN & F$\rightarrow$M & 2.20 & 2.26 \\
    (Parallel VC) & M$\rightarrow$F & 2.51 & 1.79 \\
    & {\bf AVG.} & {\bf 2.36} & {\bf 2.02} \\
    \hline
    Merlin-based & F$\rightarrow$M & 1.37 & 1.52 \\
    baseline & M$\rightarrow$F & 1.52 & 1.38 \\
    (Parallel VC) & {\bf AVG.} & {\bf 1.45} & {\bf 1.45} \\
    \hline
  \end{tabular}
\vspace{-5mm}
\end{table}

\section{Discussion}
\label{sec:limitation}
The proposed nonparallel VC method outperformed the parallel VC baseline methods for two possible reasons. One is that DTW was not conducted and no additional errors were introduced. In addition, we found some heteronyms in the training datasets. This would have further introduced matching errors for the parallel VC baseline methods but would not affect the nonparallel VC method. Another possible reason is that the proposed CycleGAN-based nonparallel VC method can use any frame pairs for training the neural network whereas the standard parallel VC method uses only aligned paired frames (obtained via DTW).

Although the learned CycleGAN is well able to convert a speaker's voice into the voice of another speaker, it is sometimes unable to strictly guarantee that the linguistic information of the converted speech is the same as that of the source speech. For example, a phoneme /$a$:/ may be converted into /I:/, and silence and voice may be exchanged in the converted speech signal in the worst case. This is because the mapping functions are not explicitly constrained to keep the linguistic information invariant between the input and output (CycleGAN only strictly constrains the linguistic information to be invariant when the input information passes through the two ``connected'' mapping functions). Therefore, the source speech sometimes might be mapped to an unexpected phone's distribution represented by a discriminator. However, we noticed that a good model that is able to keep the linguistic information can be learned when the random seed is well selected. In our implementation, the random seed was a hyper-parameter used to generate random values for model parameter initialization and training data shuffling. Therefore, it is very important to strictly constrain the converted voice linguistic information to be invariant.

\section{Conclusion and future work}
\label{sec:conclusion}
We have developed a high-quality nonparallel VC method based on a CycleGAN. We compared the proposed method with two state-of-the-art parallel VC methods, one based on a Merlin system and the other based on a GAN. In an inter-gender conversion experiment, the proposed nonparallel method performed significantly better in terms of speech quality and speaker similarity than the two parallel methods.

Future work includes developing a method for strictly constraining the linguistic information to be invariant for CycleGAN. We also plan to further improve the speech quality and speaker similarity and 
to compare our method with others using dataset of the Voice Conversion Challenge.

\label{sec:refs}

\bibliographystyle{IEEEbib}
\bibliography{refs}

\end{document}